\documentclass[letterpaper]{article}
\usepackage{aaai23}
\usepackage{times}
\usepackage{helvet}
\usepackage{courier}
\usepackage[hyphens]{url}
\usepackage{graphicx}
\usepackage{natbib}
\usepackage{caption}
\usepackage{algorithm}

\usepackage{newfloat}
\usepackage{listings}
\DeclareCaptionStyle{ruled}{labelfont=normalfont,labelsep=colon,strut=off}
\floatstyle{ruled}
\newfloat{listing}{tb}{lst}{}
\floatname{listing}{Listing}
\usepackage{wrapfig}
\usepackage{microtype}
\usepackage{graphicx}
\usepackage{mdwlist}
\usepackage{booktabs}
\usepackage{amsmath}
\usepackage{amsthm}
\usepackage{amssymb}
\usepackage{subcaption}
\usepackage{tikz}
\usetikzlibrary{arrows.meta}
\usepackage{url}

\usepackage{algpseudocode}
\usepackage{comment}

\newcommand{\system}{\emph{SPARE}}

\newcommand{\dataset}[1]{\begin{sc} #1 \end{sc} dataset}
\newcommand{\datasetonly}[1]{\begin{sc} #1 \end{sc}}
\newcommand{\bigO}{\mathcal{O}}
\newcommand{\dagt}{$\textit{DAG}_t$}

\title{SPARE: A Single-Pass Neural Model for Relational Databases}

\author {
    Benjamin Hilprecht,\textsuperscript{\rm 1}
    Kristian Kersting, \textsuperscript{\rm 1}
    Carsten Binnig \textsuperscript{\rm 1, \rm 2}
}
\affiliations {
    \textsuperscript{\rm 1} Technical University of Darmstadt\\
    \textsuperscript{\rm 2} DFKI\\
    benjamin.hilprecht@cs.tu-darmstadt.de, kersting@cs.tu-darmstadt.de, carsten.binnig@cs.tu-darmstadt.de
}

\usepackage{bibentry}

\begin{document}
	
	\maketitle
	
	\begin{abstract}
		While there has been extensive work on deep neural networks for images and text, deep learning for relational databases (RDBs) is still a rather unexplored field.
		One direction that recently gained traction is to apply Graph Neural Networks (GNNs) to RBDs. However, training GNNs on large relational databases (i.e., data stored in multiple database tables) is rather inefficient due to multiple rounds of training and potentially large and inefficient representations.
		Hence, in this paper we propose \system{}\footnote{Presented at DLG@AAAI 2023.} (\underline{S}ingle-\underline{Pa}ss \underline{Re}lational models), a new class of neural models that can be trained efficiently on RDBs while providing similar accuracies as GNNs. For enabling efficient training, different from GNNs, \system{} makes use of the fact that data in RDBs has a regular structure, which allows one to train these models in a single pass while exploiting symmetries at the same time. Our extensive empirical evaluation demonstrates that \system{} can significantly speedup both training and inference while offering competitive predictive performance over numerous baselines.
	\end{abstract}
	
\section{Introduction}

\paragraph{Motivation.}  Recent deep neural models such as aggregated embedding networks or transformers
\cite{brown2020language} have shown remarkable performance on datasets consisting of images and text.
For relational databases (RDBs), however, which are arguably a popular data source \cite{kagglerelational} in many domains, there unfortunately exists only limited work for efficiently learning neural models.
Overall, we see two main challenges for applying neural approaches to relational databases that must both be addressed: (i) First, data in such RDBs exhibits a complex (but regular) relational structure, i.e., there are multiple tables with relationships among them, which requires \textit{expressive} models. (ii) Second, RDBs are often large in size with millions or even billions of tuples and tens to hundreds of tables, which motivates the need for \textit{efficiency} of both training and inference of the models.

One direction that has recently gained traction is to apply Graph Neural Networks (GNNs) to relational databases, see e.g.~\cite{li2016gated, DBLP:conf/nips/HamiltonYL17, kipf2017semi, velickovic2018graph, xu2018how,cvitkovic2020supervised}.
Here, tuples of tables are represented as vertices and relationships between tuples of different tables (i.e., foreign keys) correspond to edges. 
In addition, to better capture relational data, it was suggested to use different weights depending on the relation types \cite{schlichtkrull2018ergcn}, specialized convolution operators \cite{ijcai2020-175} or generative architectures \cite{ijcai2019-489} on relational graphs.
While this line of work allows in principle to encode complex relational structures in an \emph{expressive} model, all these approaches treat relational data ``just'' as a graph and ignore
the more regular structures in RDBs dictated by the database schema.
Consequently, symmetries among sub-graphs are ignored and multiple rounds of message passing are required, which leads to \emph{inefficient} training and inference on large databases.

Therefore, one often follows an alternative approach in practice, namely,
to simply join the data of different tables, materialize the output of the multi-way join in one large table, and train a predictive 
model on the resulting table \cite{kanter2015deep}. 
This approach, however, not only comes with potentially high upfront costs of joining potentially many large tables but also the structure in a relational database is ignored, since all data is represented as a single flat table in the model.
To avoid the high upfront cost of joining, factorized ML approaches \cite{10.1145/2882903.2882952,10.1145/2882903.2882939} have been proposed.
While they clearly avoid expensive joins upfront, factorized approaches
conceptually still learn from a single (flat) table (i.e., the output of the join) and thus
they do not take the relational structure into account either. 
Indeed, there also exist approaches such as statistical relational learning (SRL) \cite{raedt2016statistical, natarajan2014boosted, koller2007introduction} that can better reflect the relational structure of data. However, as we show in our evaluation different from neural learning approaches, SRL typically only achieves a significantly lower accuracy. 

\textbf{Contributions.} 
To overcome the limitations, we propose a novel neural learning architecture and  procedure called \system{} tailored to the characteristics of RDBs, providing both the \emph{expressiveness} to support the schema of complex relational databases as well as an \emph{efficient} training on potentially large databases. 

Similar to GNNs, \system{} also uses a graph-based encoding for representing relational data. However, there are several important differences in how the graph encoding is constructed and how a model is learned over the graph:
(1) For \system{} we use a novel encoding using directed acyclic graphs (DAGs) that leverages the regular structure of data in RDBs, which is encoded by the schema. \system{} thus enables single-pass learning and inference. This is a major benefit over GNNs, which require multiple passes and makes the training and inference significantly more efficient. 
(2) The DAG representation in \system{} allows one to exploit the symmetries in the representation of an RDB as a set of graphs. In particular, redundant sub-graphs are systematically avoided due to relational DAG pruning. This not only reduces the size of the graphs and results in more efficient training but also reduces noise.
In our empirical evaluation, we show that \system{} is thus able to provide significantly faster training and inference while offering competitive accuracies across a wide spectrum of datasets. 

To summarize, the contributions are as follows: 
(i) We introduce single-pass learning for RDBs by transforming the data into directed acyclic graphs (DAG) and avoid repeating sub-DAGs and
(ii) we empirically show that \system{} offers significantly faster training and inference while offering competitive predictive performance. 	
\section{Problem Statement and Background}

In the following, we first discuss the general problem statement of learning a predictive model over relational databases (RDBs). Afterwards, we present how GNNs are typically being used for RDBs as a relevant background for this paper.

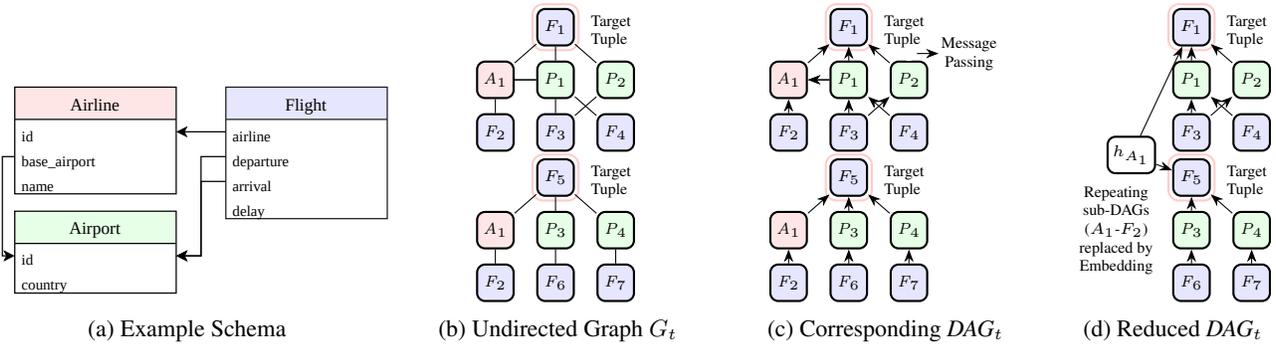
\begin{figure*}[t]
	\centering
	\subcaptionbox{Example Schema \label{fig:overview:a}}[0.3\linewidth]{\includegraphics[width=0.99\linewidth]{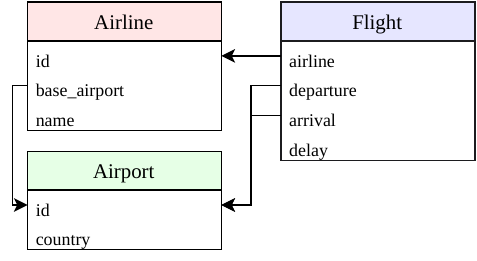}}
	\subcaptionbox{Undirected Graph $G_t$\label{fig:overview:b}}[0.24\linewidth]{
		\begin{tiny}
			\begin{tikzpicture}
			\begin{scope}[every node/.style={rounded corners=0.5em,thick,draw,minimum size=2.3em}]
			\node [fill=blue!10] (F1) at (0,0) {$F_1$};
			\node [red!20, rectangle, rounded corners, minimum size=2.9em, label={[align=left,xshift=3.5em,yshift=-3.2em] Target \\ Tuple}] (F1l) at (0,0) {};
			\node [fill=red!10] (A1) at (-0.8,-0.7) {$\mathit{A}_1$};
			\node [fill=green!10] (AP1) at (0.0,-0.7) {$\mathit{P}_1$};
			\node [fill=green!10] (AP2) at (0.8,-0.7) {$\mathit{P}_2$};
			
			\node [fill=blue!10] (F2) at (-0.8,-1.4) {$\mathit{F}_2$};
			\node [fill=blue!10] (F3) at (-0,-1.4) {$\mathit{F}_3$};
			\node [fill=blue!10] (F4) at (0.8,-1.4) {$\mathit{F}_4$};
			
			\node [fill=blue!10] (F5) at (0,-2) {$F_5$};
			\node [red!20, rectangle, rounded corners, minimum size=2.9em, label={[align=left,xshift=3.5em,yshift=-3.2em] Target \\ Tuple}] (F1l) at (0,-2) {};
			\node [fill=red!10] (A12) at (-0.8,-2.7) {$\mathit{A}_1$};
			\node [fill=green!10] (P3) at (0.0,-2.7) {$\mathit{P}_3$};
			\node [fill=green!10] (P4) at (0.8,-2.7) {$\mathit{P}_4$};
			
			\node [fill=blue!10] (F6) at (-0.8,-3.4) {$\mathit{F}_2$};
			\node [fill=blue!10] (F7) at (-0,-3.4) {$\mathit{F}_6$};
			\node [fill=blue!10] (F8) at (0.8,-3.4) {$\mathit{F}_7$};
			\end{scope}

			\begin{scope}[>={Stealth[black]},
			every node/.style={fill=white,circle},
			every edge/.style={draw=black}]
			
			\path [-] (A1) edge (F1);
			\path [-] (AP1) edge (F1);
			\path [-] (AP2) edge (F1);
			\path [-] (AP1) edge (A1);
			
			\path [-] (F3) edge (AP1);
			\path [-] (F2) edge (A1);
			\path [-] (F3) edge (AP2);
			\path [-] (F4) edge (AP1);
			
			\path [-] (A12) edge (F5);
			\path [-] (P3) edge (F5);
			\path [-] (P4) edge (F5);
			\path [-] (AP1) edge (A1);
			
			\path [-] (F6) edge (A12);
			\path [-] (F7) edge (P3);
			\path [-] (F8) edge (P4);
			\end{scope}
			\end{tikzpicture}	
		\end{tiny}
	}
	\subcaptionbox{Corresponding \dagt{}\label{fig:overview:c}}[0.24\linewidth]{
		\begin{tiny}
			\begin{tikzpicture}
			\begin{scope}[every node/.style={rounded corners=0.5em,thick,draw,minimum size=2.3em}]
			
			\node [fill=blue!10] (F1) at (0,0) {$F_1$};
			\node [red!20, rectangle, rounded corners, minimum size=2.9em, label={[align=left,xshift=3.5em,yshift=-3.2em] Target \\ Tuple}] (F1l) at (0,0) {};
			\node [fill=red!10] (A1) at (-0.8,-0.7) {$\mathit{A}_1$};
			\node [fill=green!10] (AP1) at (0.0,-0.7) {$\mathit{P}_1$};
			\node [fill=green!10] (AP2) at (0.8,-0.7) {$\mathit{P}_2$};
			
			\node [fill=blue!10] (F2) at (-0.8,-1.4) {$\mathit{F}_2$};
			\node [fill=blue!10] (F3) at (-0,-1.4) {$\mathit{F}_3$};
			\node [fill=blue!10] (F4) at (0.8,-1.4) {$\mathit{F}_4$};
			
			\node [fill=blue!10] (F5) at (0,-2) {$F_5$};
			\node [red!20, rectangle, rounded corners, minimum size=2.9em, label={[align=left,xshift=3.5em,yshift=-3.2em] Target \\ Tuple}] (F1l) at (0,-2) {};
			\node [fill=red!10] (A12) at (-0.8,-2.7) {$\mathit{A}_1$};
			\node [fill=green!10] (P3) at (0.0,-2.7) {$\mathit{P}_3$};
			\node [fill=green!10] (P4) at (0.8,-2.7) {$\mathit{P}_4$};
			
			\node [fill=blue!10] (F6) at (-0.8,-3.4) {$\mathit{F}_2$};
			\node [fill=blue!10] (F7) at (-0,-3.4) {$\mathit{F}_6$};
			\node [fill=blue!10] (F8) at (0.8,-3.4) {$\mathit{F}_7$};
			
			\end{scope}
			
			\begin{scope}[>={Stealth[black]},
			every node/.style={fill=white,circle},
			every edge/.style={draw=black}]
			
			\node [] (D1) at (0.8,-0.35) {};
			\node [] (D2) at (1.3,-0.35) {};
			\node [rectangle, minimum width=2em, minimum height=2em, align=center] (MP) at (1.6,-0.35) {Message \\Passing};
			
			\path [->] (D1) edge (D2);
			
			\path [->] (A1) edge (F1);
			\path [->] (AP1) edge (F1);
			\path [->] (AP2) edge (F1);
			\path [->] (AP1) edge (A1);
			\path [->] (F3) edge (AP1);
			\path [->] (F2) edge (A1);
			\path [->] (F3) edge (AP2);
			\path [->] (F4) edge (AP1);
			
			\path [->] (A12) edge (F5);
			\path [->] (P3) edge (F5);
			\path [->] (P4) edge (F5);
			\path [->] (AP1) edge (A1);
			\path [->] (F6) edge (A12);
			\path [->] (F7) edge (P3);
			\path [->] (F8) edge (P4);
			\end{scope}
			\end{tikzpicture}	
	\end{tiny}}
		\subcaptionbox{Reduced \dagt{}\label{fig:overview:d}}[0.18\linewidth]{
		\begin{tiny}
			\begin{tikzpicture}
			\begin{scope}[every node/.style={rounded corners=0.5em,thick,draw,minimum size=2.3em}]
			
			\node [fill=blue!10] (F1) at (0,0) {$F_1$};
			\node [red!20, rectangle, rounded corners, minimum size=2.9em, label={[align=left,xshift=3.5em,yshift=-3.2em] Target \\ Tuple}] (F1l) at (0,0) {};
			\node [] (A1) at (-0.8,-1.7) {$h_{A_1}$};
			\node [fill=green!10] (AP1) at (0.0,-0.7) {$\mathit{P}_1$};
			\node [fill=green!10] (AP2) at (0.8,-0.7) {$\mathit{P}_2$};
			
			\node [fill=blue!10] (F3) at (-0,-1.4) {$\mathit{F}_3$};
			\node [fill=blue!10] (F4) at (0.8,-1.4) {$\mathit{F}_4$};
			
			\node [fill=blue!10] (F5) at (0,-2) {$F_5$};
			\node [red!20, rectangle, rounded corners, minimum size=2.9em, label={[align=left,xshift=3.5em,yshift=-3.2em] Target \\ Tuple}] (F1l) at (0,-2) {};
			\node [fill=green!10] (P3) at (0.0,-2.7) {$\mathit{P}_3$};
			\node [fill=green!10] (P4) at (0.8,-2.7) {$\mathit{P}_4$};
			
			\node [fill=blue!10] (F7) at (-0,-3.4) {$\mathit{F}_6$};
			\node [fill=blue!10] (F8) at (0.8,-3.4) {$\mathit{F}_7$};

			\end{scope}
			
			\begin{scope}[>={Stealth[black]},
			every node/.style={fill=white,circle},
			every edge/.style={draw=black}]
			
			\node [rectangle,minimum width=2.5em, minimum height=2em, align=center] (Emb) at (-1,-2.7) {Repeating \\sub-DAGs \\($A_1$-$F_2$) \\ replaced by\\ Embedding	
		};
			
			\path [->] (A1) edge (F1);
			\path [->] (AP1) edge (F1);
			\path [->] (AP2) edge (F1);
			\path [->] (F3) edge (AP1);
			\path [->] (F3) edge (AP2);
			\path [->] (F4) edge (AP1);
			
			\path [->] (A1) edge (F5);
			\path [->] (P3) edge (F5);
			\path [->] (P4) edge (F5);
			\path [->] (F7) edge (P3);
			\path [->] (F8) edge (P4);
			\end{scope}
			\end{tikzpicture}	
	\end{tiny}}
	\vspace{-0.5ex}
	\caption{Example of \system{} on a database with three tables (a). GNNs use undirected graphs as shown in (b). In the example, $G_t$ consists of arrival and departure airports, airline and corresponding flight tuples. \system{} instead uses a directed acyclic graph \dagt{} as shown in (c), which defines the message passing order. For reducing the size of graph \dagt{}, \system{} leverages relational DAG pruning which replaces sub-DAGs with an embedding as shown in (d).}
	\label{fig:overview}
\end{figure*}

\textbf{Problem Statement.}
Supervised learning on RDBs aims at predicting the value of an attribute (called target attribute $a_t$) of a given tuple $t$ in a relational database. 
As an example, in a flights database such as the one shown in in Fig. \ref{fig:overview:a}, we might want to learn a model that allows us to predict the delay which is an attribute of a flight tuple given the information from the flight and from the airline and airport tables.

More precisely, we define an RDB as a set of tables $T_i$ where each table $T_i$ with attributes $A_{i1},\dots,A_{in}$ consists of tuples $t_i\in T_i$ of the form $t_i=(a_{i1},\dots, a_{in})$.
Tuples of different tables can be connected via foreign key relationships, e.g., a flight refers to a departure and arrival airport.
For learning a model, one of the tables in the RDB is the target table $T_t$ for which we want to predict one of the attributes $a_{t}$, called the target attribute.

\textbf{GNNs for Supervised Learning on RDBs.}
Recently, it was shown that GNNs can accurately capture the relational structure in RDBs and are thus beneficial for supervised learning on RDBs \cite{cvitkovic2020supervised}. For every target tuple $t,$ for which we want to predict the target attribute $a_t$, we can construct a graph representing the relational structure of related tuples, apply a GNN on this graph and finally predict the target attribute using the GNN. More precisely, the graph contains tuples as nodes, which are connected via foreign key relationships (typically connected tuples are limited to a certain depth/number of hops to limit the size of the graphs). These relationships then correspond to edges in the graph. For instance, if we want to predict the delay of a flight, we create a separate undirected graph for every individual flights tuple which would be the target tuple of a graph (cf. Fig. \ref{fig:overview:b}).

One could wonder why it is required to construct one graph per target tuple instead of representing the entire RDB as a single large graph which could improve the performance of learning and avoid redundant copies of tuples in different sub-graphs. However, this would result in transductive learning \cite{DBLP:conf/nips/HamiltonYL17} and thus the learned model would not generalize to unseen tuples, which can frequently occur due to database updates.

\textbf{Training Procedure of GNNs.} 
The learning of GNNs (in particular those used for supervised learning on RDBs) is typically based on the message passing paradigm \cite{pmlr-v70-gilmer17a}. That is, each vertex $v$ of a graph representation is initialized with a hidden state vector $h_v^0$ and then updated in $T$ rounds of message passing. 
Specifically, per message passing round $t$, each vertex $v$ sends a message $m_{vw}^t$ to each of its neighbors $w$. The message is computed using a learnable function taking optional edge features as well as the hidden states of the vertex $h_v^t$ and of its neighbors $h_w^t$ of round $t$ as input.
Then, each vertex aggregates its incoming messages using a learnable function and updates the hidden state to obtain $h_v^{t+1}$. After the message passing, a readout function aggregates all hidden states $h_v^{T}$ from all vertexes to compute the final prediction for the graph. 
Hence, for $T$ rounds of message passing we need to compute $\bigO(|E|T)$ messages and update $\bigO(|V|T)$ hidden states, where $|V|$ and $|E|$ denote the number of vertices and edges, respectively. 

\section{\system{} -- Single-Pass Relational Models}
\label{sec:learning}

We now introduce \system{}, which provides efficient and accurate supervised learning on RDBs.
The core idea is to represent related tuples in a DAG instead of an undirected graph, which then enables a single-pass message passing training and inference. In particular, the edge directions in the DAG dictate how messages are propagated in the graph s.t. a single bottom-up pass in the DAG is sufficient to encode the entire graph. 

In the remainder of this section, we show how to construct a \dagt{} from a multi-relational database. We then provide details of our single-pass learning process on this model architecture and introduce an important optimization that significantly reduces the size of the graph \dagt{} by leveraging the relational structure of the data.

\subsection{DAG Construction}
For the sake of better understanding,
we assume an undirected graph $G_t$ as constructed for GNNs is given, which is then transformed into a directed graph \dagt{} for \system{}.
This is illustrated in Fig.~\ref{fig:overview:c}. It shows the directed graphs of tuples for two target flight tuples $F_1$ and $F_5$ that are constructed by transforming the undirected graphs in Fig.~\ref{fig:overview:b}. 
However, we would like to stress that the directed graphs \dagt{} could clearly also be constructed directly from a relational database.

To transform $G_t$ into a directed acyclic graph \dagt{}, the idea of \system{} is to perform a breadth-first traversal (BFT) on $G_t$, starting from the target tuple $t$. When traversing the graph using BFT we introduce an edge in \dagt{} if the source vertex in $G_t$ has a higher depth than the target vertex. More technically, the BFT assigns a depth $d(v)$ to every vertex $v$. We introduce a directed edge $(v,w)$ in \dagt{} if there is an undirected edge between $v$ and $w$ in $G_t$ and $d(v)>d(w)$.
Unfortunately, this transformation can still lead to edges in $G_t$ not having a counterpart in \dagt{}. 
This is the case if two vertices $v$ and $w$ have the same depth in $G_t$. 
In our running example this would happen for the airline tuple $A_1$ and airport tuple $P_1$ in the graph in Fig.~\ref{fig:overview:b}.
To encode such relationships between tuples of the same depth, we define a fixed (but arbitrary) global ordering between all tuples. Using this order, we add a directed edge $(v,w)$ to \dagt{} if $d(v)=d(w)$ and the corresponding tuple of $v$ has a higher order than the one of $w$. It can easily be shown that the resulting graph is a DAG where the root node corresponds to the target tuple.

\subsection{Learning Procedure}

In the following, we describe the algorithm used for single-pass (i.e., the bottom-up) forward propagation in a \system{} model, i.e., assuming that the parameters have already been learned. The learning of the parameters can happen end-to-end since all steps in our algorithm are differentiable. 
The pseudo code of the bottom-up propagation algorithm is given in Algorithm \ref{alg:forward_prop}.

Recall that \system{} models operate on a \dagt{} to learn a predictive model for a given target attribute $a_t$. As a first step of the forward pass, we thus compute a hidden representation $h_v^0$ for every vertex in \dagt{} that represents an embedding for every tuple similar to GNNs. This is done by encoding the attribute (feature) vector $x_v$ of the corresponding tuple $t_v$ and then feeding these into a table-specific MLP such that we obtain a fixed-sized vector for every tuple in the graph. 

We then propagate the information through the \dagt{} using a single pass by traversing the DAG in a topological ordering.
During this traversal, we compute the hidden state of every vertex after message propagation using the child message aggregation function $M$. 
As a result of the bottom-up pass, we thus obtain a hidden representation $h_t$ of the target tuple $t$, which is then fed into a final MLP to obtain the prediction.

\begin{algorithm}[t]
	\small
	\caption{\system{} Single-Pass Propagation Algorithm (Forward Pass)}
	\label{alg:forward_prop}
	\begin{algorithmic} 
		\State \textbf{Input:} DAG of target tuple \dagt{}, input features per tuple $\{x_v \mid v\in \mathit{DAG}_t\},$ table-specific encoders $\mathit{MLP}_T$, children aggregation function $M$, final output function $\mathit{MLP}_{\mathit{out}}$  
		\State \textbf{Output: } Prediction for target tuple $h_t$ 
		\State
		
		\For{$T$ in Tables}  \Comment{Encode tuple features with table-specific MLP}
		\For{$v\in$\dagt{} belonging to tuples in $T$}
			\State $h_v^0 \leftarrow \mathit{MLP}_{T}(x_v)$
		\EndFor
		\EndFor
		
		\For{$v\in$\dagt{} in topological ordering} \Comment{Single bottom-up pass in DAG}
			\State $h_v \leftarrow M(\{h_w \mid w \in \mathit{children}(v)\}, h_v^0)$ \Comment{Parallelism by batching independent nodes} \label{line:child_agg}
		\EndFor
		\State \Return $\mathit{MLP}_{\mathit{out}} (h_t)$
	\end{algorithmic}
\end{algorithm}

We have not yet specified the architecture used for the child aggregation function $M$, which combines the hidden states of the child nodes (i.e., the predecessor nodes in topological ordering). 
While it is possible to leverage architectures tailored for sets such as DeepSets \cite{NIPS2017_f22e4747} or Set Transformers \cite {pmlr-v97-lee19d}, using graph kernels such as GAT \cite{velickovic2018graph} or GCN \cite{kipf2017semi} worked best in our experiments.

\textbf{Discussion.} Overall, the single-pass learning of \system{} requires $\bigO(|E|)$ messages and  $\bigO(|V|)$ updates of hidden states instead of $\bigO(|E|T)$ messages and $\bigO(|V|T)$ updates which significantly reduces the runtime complexity. 
One could now think that compared to undirected message passing in GNNs, the directed (bottom-up) message passing of \system{} results in an inferior parallelism since the computation has to be executed in the topological ordering of the DAG. While the minimum number of sequential operations on a DAG thus coincides with the longest directed path on a DAG, this does not result in an inferior performance since \system{} offers parallelism by (i) computing the message passing for independent nodes in a single DAG in parallel and (ii) batching several graphs as in traditional GNN processing. 

\subsection{Relational DAG Pruning}
We now show that \system{} can be further improved using relational DAG pruning which builds upon the specific properties of relational schemas. The main idea of this optimization is based on the observation that sub-DAGs often occur repeatedly in DAGs of different target tuples during training and can therefore be replaced by a single node that represents the embedding for a sub-DAG.
Doing so can better reflect symmetries in the data and thus reduces the size of the graphs and enables more efficient training.

More technically, during training some subgraphs repeatedly appear in the \dagt{} of different target tuples.
To reflect this symmetry, \system{} replaces redundant sub-graphs with a learnable embedding $h_e$. 
The embedding itself is in fact just a hidden state $h_v$ that can be trained end-to-end similar to all other parameters in a \system{} model. However, by using an embedding instead of a sub-DAG, we avoid the effort of actually computing the repeated bottom-up pass on the sub-DAG for every target tuple. 
Consider the flight delay prediction example in Fig.~\ref{fig:overview:c}. Here, a limited number of airline vertices (e.g., $A_1$) will appear in the \dagt{} of different target flight tuples.
In particular, the corresponding sub-DAGs ($A_1$ and $F_2$) that represent an airline and the flights will be similar for the different \dagt{}'s of different flight target tuples and could thus be replaced by an embedding.

An important question is how redundant sub-DAGs can be recognized efficiently. Due to the way how we construct the DAGs, we can be sure that if we encounter a certain tuple during the traversal (at the same depth) multiple times, the sub-DAG below this tuple will be the same. Hence, it is sufficient to keep track of how often tuples are encountered at certain depths during the traversal. If a certain (tunable) threshold is exceeded, we can thus replace the sub-DAG following the tuple with an embedding. Similarly, we can just stop the traversal at tuples of particularly small tables.

\section{Empirical Evaluation}
\label{sec:empirical_evaluation}

Due to the large data sizes present in relational databases (RDBs), performance is key to any approach operating on RDBs. We thus investigate whether single-pass models accelerate learning and inference compared to other neural architectures for supervised learning on RDBs such as GNNs. In addition, we evaluate whether potential speedups come at the cost of a reduced accuracy. Overall, we find that \system{} offers a competitive predictive performance while more importantly accelerating training and inference up to 9.7x and 5x compared to RGCNs and GCNs, respectively.

\subsection{Training and Inference Performance}
\label{sec:performance}

We first compare the training and inference performance of \system{} with neural architectures for supervised learning on RDBs. We evaluate Spare on five different datasets: \datasetonly{Airline} \cite{airline_dataset}, \datasetonly{Airbnb} \cite{airbnb_dataset}, \datasetonly{Baseball} \cite{baseball_dataset}, \datasetonly{Credit Risk} \cite{homecreditdefaultrisk_dataset}, and \datasetonly{IMDB} \cite{imdb_dataset}, which are widely used for evaluating supervised learning tasks on databases where one categorical or continuous attribute should be predicted. The datasets are of varying complexity between 7 and 25 tables with up to 36 relationships among tables, and between about 0.5M up to 71.4M tuples which reflects the requirement of an efficient learning.

\textbf{Experimental Setup.} We compare the performance of \system{} with five baselines. First, we employ the popular GAT \cite{velickovic2018graph} and GCN \cite{kipf2017semi} graph kernels. In particular, we apply message passing on the graph of related tuples $G_t$ and afterwards use a readout function such that all tuples of the graph are considered. In addition, we evaluate GraphSAGE \cite{DBLP:conf/nips/HamiltonYL17} where only the node embedding of the target node is computed and thus the message passing overhead is reduced. We moreover evaluate the relational variants \cite{schlichtkrull2018ergcn} called RGAT and RGCN that use different parameters depending on the edge type (i.e., which tuple relationship).\footnote{However, since for relational GNNs we experienced a significant performance degradation (10x on some datasets) and exceeded the GPU memory in many cases, we instead concatenate the input hidden states with a learnable embedding based on the relation types.} Note that we include additional baselines in the evaluation of the predictive performance. However, since these resulted in significantly inferior accuracies, we restrict ourselves to neural baselines for the evaluation of the performance since we are aiming at both efficient and expressive models.

To exclude the effect of using different hyperparameters on the performance, we chose identical hyperparameters for \system{} and GNNs when applicable (e.g., sizes of hidden layers etc.)  To ensure a fair comparison we set the number of message passing rounds in all GNNs to just $T=2$ and for GraphSAGE to the depth of the graph $G_t$ (s.t. all tuples can be taken into account for the prediction). 

\begin{figure*}[t]
	\begin{center}
		\centerline{\includegraphics[width=0.9\linewidth]{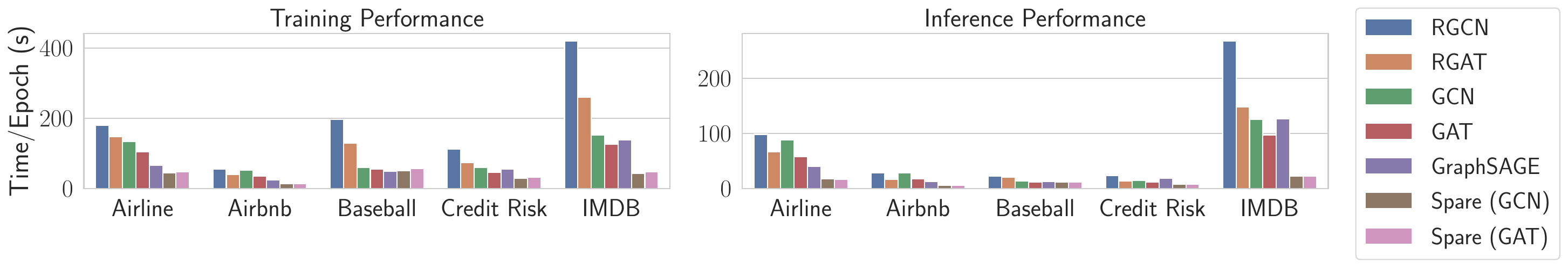}}
		\vspace{-1.5ex}
		\caption{Training and inference times per epoch. \system{} offers significant speedups which are crucial for large data sizes typical for RDBs due to (i) single-pass message passing which reduces the number of passed messages (from $\bigO(|E|*T)$ to $\bigO(|E|)$ for $T$ rounds of message passing) and (ii) reduced graph sizes due to relational DAG pruning.}
		\vspace{-4.5ex}
		\label{fig:training-speedup}
	\end{center}
\end{figure*}

\begin{table*}[t]
	\caption{Mean and standard deviation of the normalized RMSE and AUROC for regression and datasets, respectively for baselines and \system{} over 5-fold cross validation. The best ("$\bullet$") and runner-up ("$\circ$") results per dataset are bold. Overall, \system{} shows competitive predictive performance while offering significant speedups during training and inference.}
	\label{tab:predictive-performance}
	\begin{center}
		\begin{small}
\begin{tabular}{lrrrrr} 
\toprule 
 & Airline & Airbnb & Baseball & Credit Risk & IMDB\\ 
 & (NRMSE) & (NRMSE) & (NRMSE) & (AUROC) & (AUROC)\\ 
 \midrule 
Single Table GBDT & 0.163$\pm$0.002 & 0.970$\pm$0.001 & 0.915$\pm$0.004 & 0.756$\pm$0.002 & 0.773$\pm$0.007 \\ 
BoostSRL & 1.379$\pm$0.007 & 1.042$\pm$0.001 & 1.254$\pm$0.015 & 0.712$\pm$0.015 & 0.503$\pm$0.009 \\ 
DFS GBDT & 0.120$\pm$0.001 & 0.945$\pm$0.001 & 0.743$\pm$0.006 & 0.776$\pm$0.003 & 0.883$\pm$0.002 \\ 
GNN (GCN) & 0.057$\pm$0.003 & 0.928$\pm$0.001 & 0.524$\pm$0.011 & 0.777$\pm$0.003 & 0.910$\pm$0.006 \\ 
GNN (GAT) & 0.243$\pm$0.101 & 0.929$\pm$0.001 & 0.504$\pm$0.004 & 0.777$\pm$0.004 & 0.906$\pm$0.006 \\ 
RGNN (RGCN) & 0.046$\pm$0.003 & $\bullet$\textbf{0.926$\pm$0.001} & 0.534$\pm$0.007 & 0.779$\pm$0.002 & 0.914$\pm$0.000 \\ 
RGNN (RGAT) & 0.249$\pm$0.018 & 0.928$\pm$0.001 & 0.503$\pm$0.013 & $\circ$\textbf{0.779$\pm$0.003} & 0.911$\pm$0.003 \\ 
GraphSAGE & 0.052$\pm$0.001 & 0.928$\pm$0.002 & 0.506$\pm$0.011 & 0.774$\pm$0.003 & 0.917$\pm$0.001 \\ 
 \midrule
Spare (GCN) & $\circ$\textbf{0.026$\pm$0.004} & 0.928$\pm$0.002 & $\circ$\textbf{0.499$\pm$0.011} & $\bullet$\textbf{0.781$\pm$0.002} & $\circ$\textbf{0.921$\pm$0.001} \\ 
Spare (GAT) & $\bullet$\textbf{0.023$\pm$0.003} & $\circ$\textbf{0.927$\pm$0.001} & $\bullet$\textbf{0.495$\pm$0.014} & 0.778$\pm$0.003 & $\bullet$\textbf{0.921$\pm$0.001} \\ 
 \bottomrule
 \end{tabular} 			
		\end{small}
	\end{center}
	\vspace{-2.5ex}
\end{table*}

\textbf{\system{} models significantly speed up the training.} As depicted in Fig.~\ref{fig:training-speedup}, \system{} exhibits significantly more efficient training and inference with speedups of up to 9.7x, 5x and 3.2x compared to RGCNs, GCNs and GraphSAGE, respectively.  
These speedups are caused by (i) single-pass message passing and (ii) relational DAG pruning. In the following, we summarize the findings of both effects.

In particular, due to (i), we only require $|E|$ passed messages and $|V|$ updates on a graph $G(V,E)$ instead of $T\cdot|E|$ and $T\cdot|V|$, respectively, for a GNN with $T$ rounds of message passing. However, this absolute reduction in passed messages would not result in speedups if it came at the cost of decreased parallelism as discussed before. Note that the experiments demonstrate that this is not the case for \system{} since we can combine the message passing of independent nodes in a DAG and batch several DAGs to achieve higher parallelism. 
Due to (ii) we can also reduce the number of nodes $|V|$ and edges $|E|$ compared to GNNs. The magnitude of this reduction depends on the specific dataset and can be as much as 78\ for \datasetonly{IMDB} and 38\ on average. 

While GraphSAGE is more efficient than standard GNNs, SPARE still results in significant speedups up to 3.2x for two main reasons: (i) in the computation of GraphSAGE one tuple can appear more than one time (in case of cycles as opposed to SPARE where the DAG is acyclic) and (ii) GraphSAGE does not employ DAG pruning.

In addition, while for the \dataset{Baseball} we observe a similar speedup of \system{} compared to RGCNs and RGATs, the performance compared to the non-relational variants is just on par. 
A closer investigation reveals that the GPU is underutilized due to a small batch size (which is limited by the available GPU memory) and for different configurations, speedups could be observed.

\subsection{Predictive Performance}

We now evaluate the predictive performance of \system{}. To this end, we first tuned the hyperparameters for all baselines as well as the \system{} models and afterwards performed a 5-fold cross validation. Overall, we found that \system{} models outperform the baselines on four out of five datasets while being on par for the remaining dataset.

\textbf{Experimental Setup.} In addition to the GNN baselines discussed before, we include further baselines. We evaluate a popular rule-based approach to extract features from a relational dataset: deep feature synthesis (DFS) \cite{kanter2015deep} which we combine with gradient boosted decision trees (GBDT)\footnote{For GBDT models we used LightGBM \cite{NIPS2017_6449f44a}}, which has exhibited competitive performance on tabular data. To demonstrate that it is important to incorporate the relation structure, we also include GBDT models trained only on the target table (i.e., ignoring additional tables in the schema). We also evaluate boosted statistical relational learners, namely BoostSRL \cite{natarajan2014boosted,boostsrl}.

We tuned the hyperparameters of all baselines and \system{} models using the validation set. In order to alleviate the effect of different dataset sizes we sampled at most 100K tuples per epoch. For all datasets, we used a 65/15/20 split.

\textbf{\system{} exhibits competitive predictive performance.} We report the mean and standard deviation of the test normalized rooted mean squared error (NRMSE) for regression datasets (normalized using the standard deviation) and the area under curve metric (AUROC) for classification tasks in Tab. \ref{tab:predictive-performance}. 
As we can see, \system{} models are at least on par with state-of-the-art approaches and competitive for the majority of the evaluated datasets. This confirms that our approach does not trade a training speedup for lower accuracy.

Importantly, we observe that models operating on features generated by DFS already show an improved performance compared to the single table approach. However, they are still inferior to architectures that explicitly encode the structure as graphs showing that it is overall beneficial to encode the relational information in a graph-based model.
\section{Conclusion and Future Work}

We have introduced \system{} for supervised learning on relational databases (RDBs). A central problem is to scale neural architectures for the large data volumes that can often be found in RDBs. We addressed this problem by operating on DAGs instead of undirected graphs which enables a more efficient message passing and DAG pruning. We could confirm that \system{} offers significant speedups and a competitive predictive performance. 
It is thus an interesting avenue for future work to apply \system{} to graph problems where vertices exhibit similarly rich features as in RDBs and reoccurring sub-structures can be exploited.

\section{Acknowledgments}

We thank the reviewers for their feedback. This research is funded by the BMBF project KompAKI (02L10C150), the BMBF AI competence center (01IS22091), the Hochtief project \emph{AICO} (AI in Construction), as well as the HMWK cluster project \emph{3AI} (The Third Wave of AI). Finally, we want to thank hessian.AI as well as DFKI Darmstadt. 
 	
	\bibliography{literature}
	
\end{document}